\documentclass[twocolumn]{webofc}
\usepackage[varg]{txfonts}   
\woctitle{MENU 2013}
\begin{document}
\title{SAID Analysis of Meson Photoproduction: Determination of Neutron 
	and Proton EM Couplings}
%
%

\author{Igor Strakovsky\inst{1}\fnsep\thanks{\email{igor@gwu.edu}} \and
        William Briscoe\inst{1} \and
	Alexander Kudryavtsev\inst{2,1} \and
	Vladimir Tarasov\inst{2} \and
	Ron Workman\inst{1}
}
\institute{Institute for Nuclear Studies, Department of Physics, 
	The George Washington University, Washington, DC 20052, USA.
\and
           Institute of Theoretical and Experimental Physics, Moscow, 
	117259 Russia.
          }

\abstract{%
We present an overview of the GW SAID group effort to analyze 
on new pion photoproduction on both proton- and neutron-targets. 
The main database contribution came from the recent CLAS and
MAMI unpolarized and polarized measurements.  The differential 
cross section for the processes $\gamma n\to\pi^-p$ was 
extracted from new measurements accounting for Fermi motion 
effects in the impulse approximation (IA) as well as NN- and 
$\pi$N effects beyond the IA.  The electromagnetic coupling 
results are compared to other recent studies.
}
\maketitle 

\underline{SAID for Baryon Spectroscopy}. The properties of 
the resonances for the non-strange sector have been determined 
almost entirely from the results of $\pi$N elastic scattering 
analyses~\cite{PDG}. Meson photoproduction reactions have 
mainly served to fix electromagnetic (EM) couplings. With the 
refinement of multichannel fits and the availability of 
highprecision photoproduction data for both single- and 
double-meson production, identifications of some new states 
have emerged mainly due to evidence from reactions not 
involving single-pion-nucleon initial or final states~\cite{PDG}.  
The GW SAID N$^\ast$ program consists of $\pi N\to\pi N$, 
$\gamma N\to\pi N$, and $\gamma^\ast p\to\pi N$ components as 
was established by Dick Arndt on 1997.  Assuming dominance of 
two hadronic channels [$\pi$N elastic and $\pi N\to\eta N$], 
we parametrize $\gamma^\ast p\to\pi N$ in terms of $\pi N\to\pi 
N$ amplitudes (\cite{cm12} and references therein).  Most of 
the pion photoproduction analyses use SAID $\pi$N partial-wave 
analysis (PWA) outcome~\cite{piN} or its modification as input 
for the constraint as well.  However, beyond $\pi$N elastic 
scattering, single-pion photoproduction remains the most 
studied source of resonance information.  Much of the effort 
aimed at providing complete or nearly complete information for 
meson-nucleon photoproduction reactions has been directed to 
measuring double-polarization observables. However, often 
overlooked is that the data coverage for several 
single-polarization observables, also vital in determining the 
properties of the nucleon resonance spectrum, still remains 
incomplete. 

Here we focus on the single-pion production data and note 
that a complete solution requires couplings from both charged 
and neutral resonances~\cite{Wat,Wal}, the latter requiring 
$\pi^-p$ and $\pi^0n$ photoproduction off a neutron target, 
typically a neutron bound in a deuteron target. Extraction of 
the two-body ($\gamma n\to\pi^-p$ and $\gamma n\to\pi^0n$) 
cross sections requires the use of a model-dependent nuclear 
correction, which mainly comes from final-state interactions 
(FSI)~\cite{MW}.  As a result, our knowledge of the neutral 
resonance couplings is less precise than that of the charged 
values for well-known low-laying baryons. The uncertainties 
for such kind of neutral states with J$^P$ = $\frac{1}{2}$, 
for instance, N(1440)1/2$^+$, N(1535)1/2$^-$, and 
N(1650)1/2$^-$ vary from 25\% to 140\%~\cite{PDG}.  Some of 
the N$^\ast$ baryons [N(1675)5/2$^-$, for instance] have 
stronger EM couplings to the neutron than to the proton, but 
parameters are very uncertain ($N^\ast\to\gamma p: +0.019\pm 
0.008~GeV^{-1/2}$ while $N^\ast\to\gamma n: -0.043\pm 
0.012~GeV^{-1/2}$~\cite{PDG}).  Then, PDG12 estimates for 
the $A_{1/2}$ and $A_{3/2}$ proton decay amplitudes of the 
N(1720)3/2$^+$ state are consistent with zero, while the 
recent SAID determination~\cite{cm12} gives small but 
non-vanishing values.  Other unresolved issues relate to 
the second P$_{11}$, N(1710)1/2$^+$, that we do not seen in 
the recent SAID $\pi$N PWA~\cite{piN} contrary to the findings 
of other PWAs referenced by PDG12~\cite{PDG}.

\underline{Pion photoproduction off the proton}. The overall 
SAID $\chi^2$ has remained stable ($\chi^2$/data = 2.1) 
against the growing database, which has increased by a factor 
of 2 since 1995 (13.4k up to 27.3k data points)~\cite{SAID}. 
Most of this increase coming from photon-tagging facilities. 
More complete data sets for double- and single-polarization 
observables for pion photoproduction can offer important 
constraints on analyses of the photoproduction reaction. 

\begin{figure}[h]
\centerline{
\includegraphics[height=4cm, angle=90]{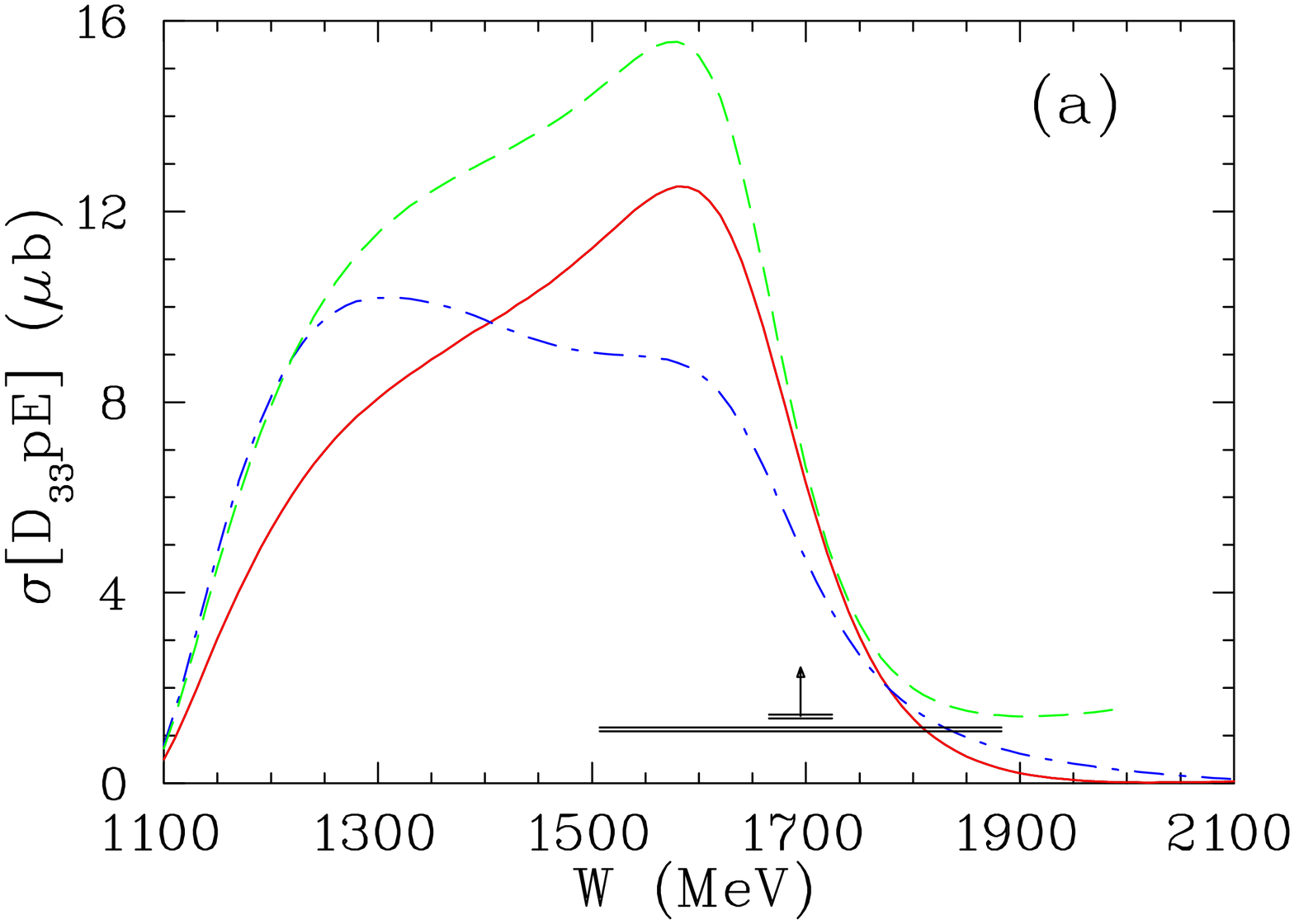}\hfill
\includegraphics[height=4cm, angle=90]{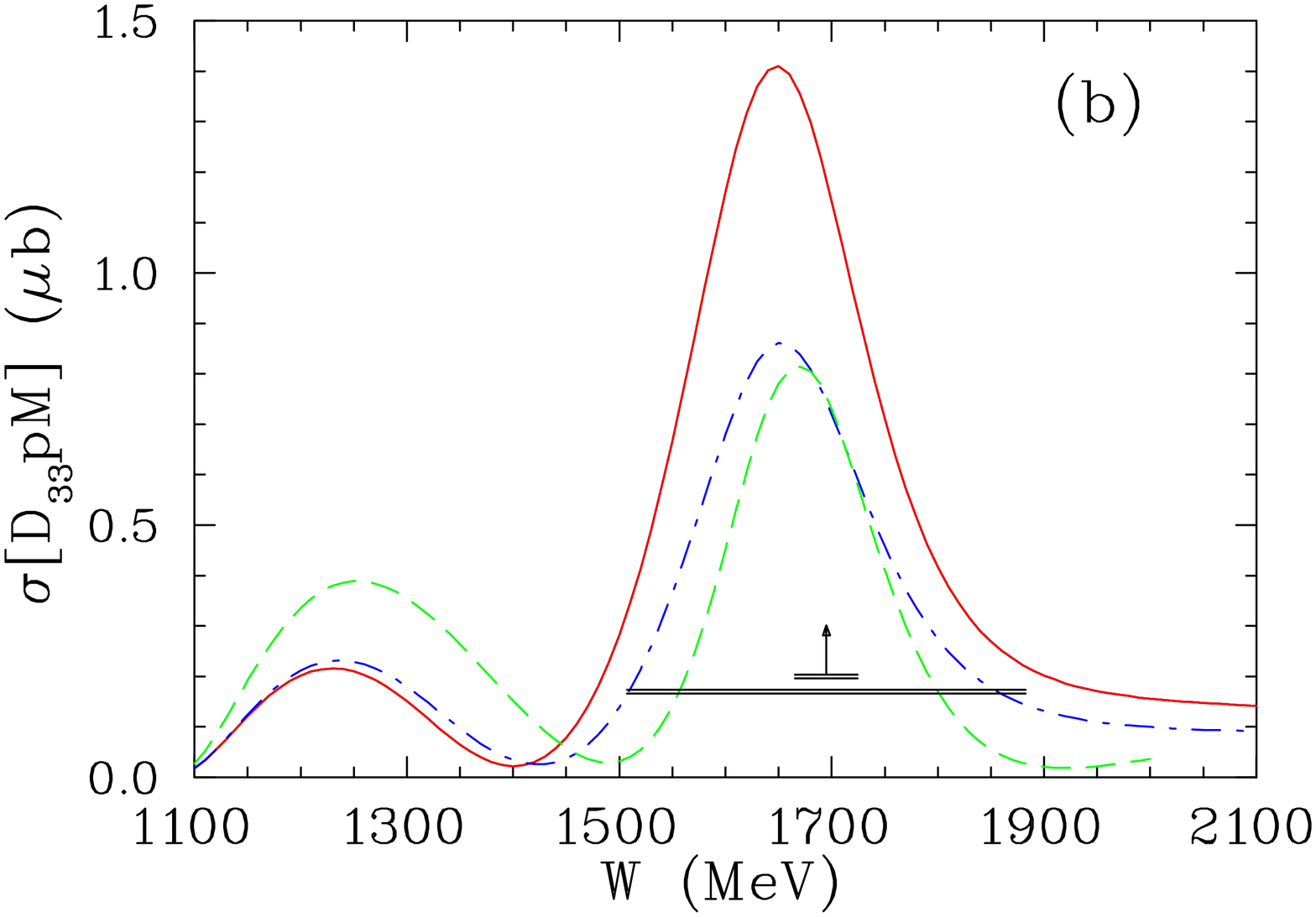}}
\centerline{
\includegraphics[height=4cm, angle=90]{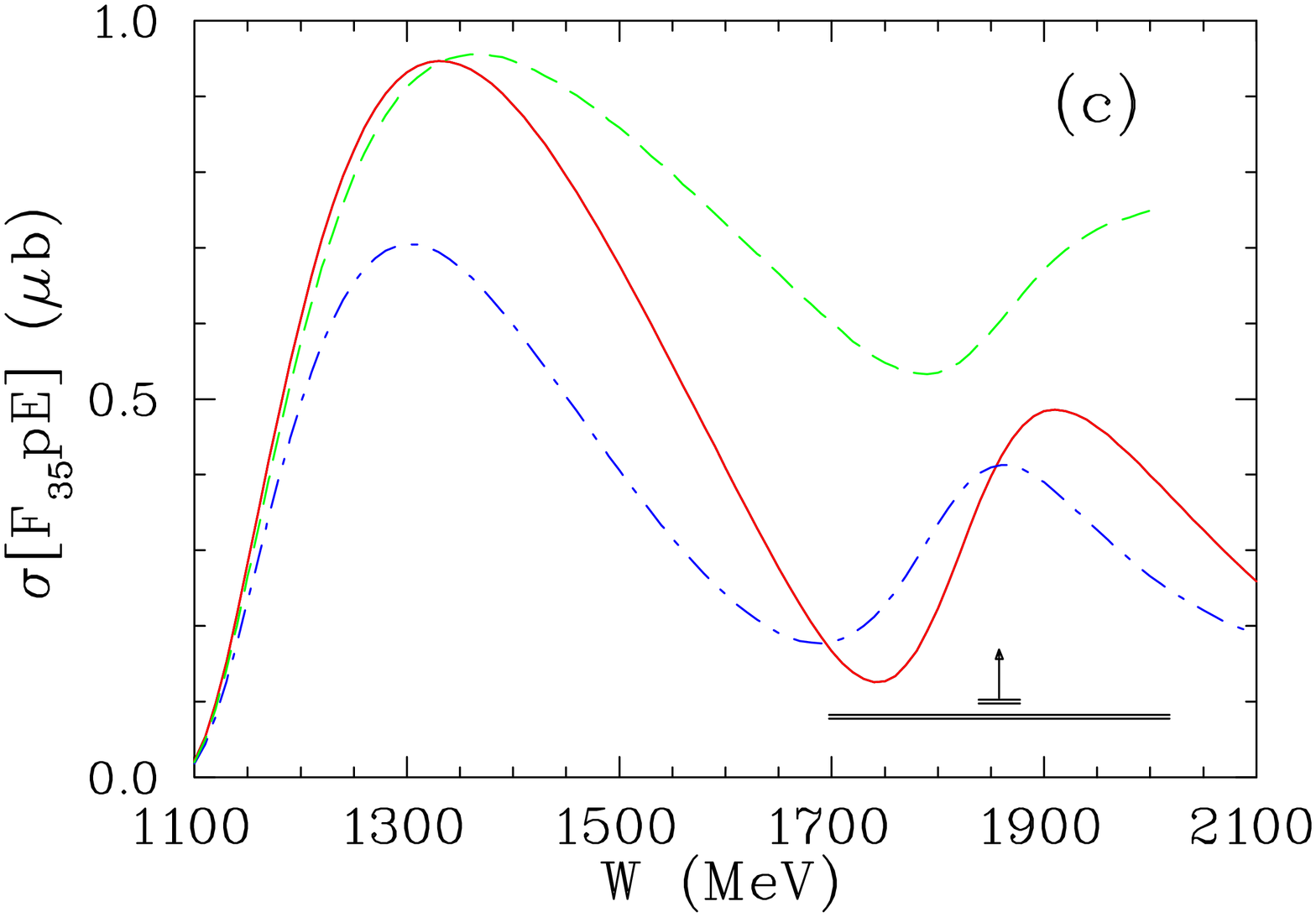}\hfill
\includegraphics[height=4cm, angle=90]{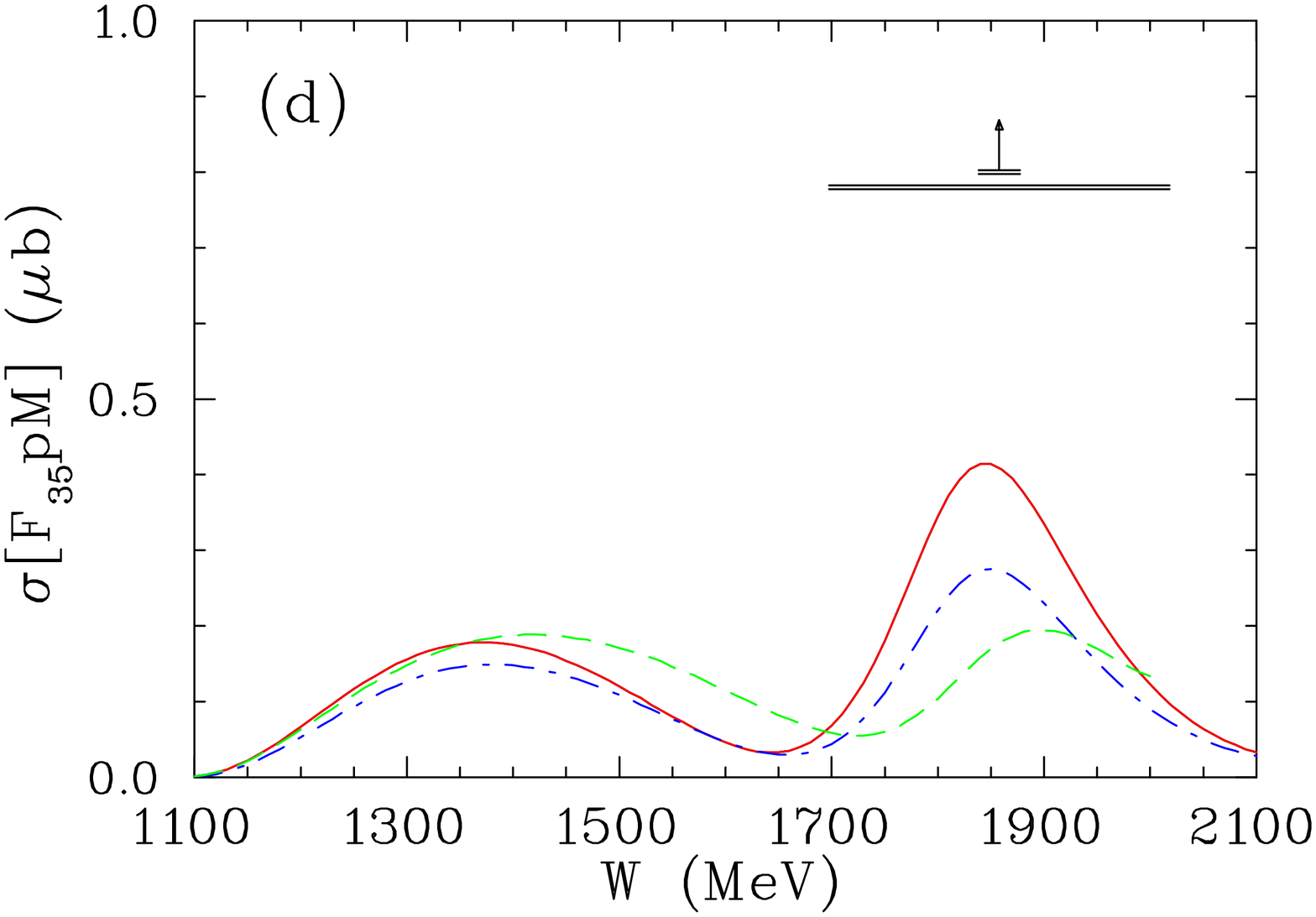}}
\caption{Partial cross sections for multipoles
	with the largest change was found after including
	the new CLAS data in the fit,
	$\Delta$(1700)3/2$^-$ and $\Delta$(1905)5/2$^+$. 
	Solid (dash-dotted) lines correspond to the SAID 
	DU13~\protect\cite{du13} (CM12~\protect\cite{cm12}) 
	solution. Dashed lines give MAID07~\protect\cite{MAID}, 
	Vertical arrows indicate resonance energies 
	W$_R$ and horizontal bars show full $\Gamma_{\pi 
	N}$ widths associated with the SAID $\pi$N
	solution WI08~\protect\cite{piN}.  \label{fig-1}}
\end{figure}

Using linearly polarized photons and an unpolarized target, 
CLAS provides a large set of beam asymmetry $\Sigma$ 
measurements for $\gamma p\to\pi^0p$ and $\gamma p\to\pi^+n$ 
from E$_\gamma$ = 1.100 and up to 1.860~GeV in laboratory 
photon energy, corresponding to a CM energy W range of 1.7 
$-$ 2.1~GeV ($\theta$ = 30 $-$ 150$^\circ$ of pion production 
angle in CM)~\cite{du13}.  Its contribution to the world 
database is more than doubled~\cite{SAID}.  In Figs.~\ref{fig-1}, 
we show the effect of new CLAS $\Sigma$ measurements in terms 
of partial cross sections from SAID (CM12~\cite{cm12} and 
recent DU13~\cite{du13} included new CLAS data) and MAID
\cite{MAID}. While the CM12 and DU13 solutions differ over 
the energy range of the recent CLAS experiment, the resonance 
couplings are fairly stable. The largest change is found for 
the $\Delta$(1700)3/2$^-$ and $\Delta$(1905)5/2$^+$ states, 
for which the various analyses disagree significantly in terms 
of photo-decay amplitudes (Table~\ref{tbl-1}).  
\begin{table}
\centering
\caption{Proton helicity amplitudess $pA_{1/2}$ and $pA_{3/2}$
        (in [(GeV)$^{-1/2} \times 10^{-3}$] units).}
\vspace{2mm}
{\begin{tabular}{|c|c|c|c|}
\hline
Resonance           & $pA_{1/2}$ &  $pA_{3/2}$ & Ref.\\
\hline
$\Delta(1700)3/2^-$ & 132$\pm$ 5 &  108$\pm$ 5 &SAID~DU13~\protect\cite{du13}\\
		    & 105$\pm$ 5 &   92$\pm$ 4 &SAID~CM12~\protect\cite{cm12}\\
                    & 160$\pm$20 &  165$\pm$25 &BnGa12~\protect\cite{bnga12}\\
                    &  58$\pm$10 &   97$\pm$ 8 &Kent12~\protect\cite{kent}\\
                    & 226        &  210        &MAID~\protect\cite{MAID}\\
                    & 104$\pm$15 &   85$\pm$22 &PDG12~\protect\cite{PDG}\\
\hline
$\Delta(1905)5/2^+$ &  20$\pm$ 2 &$-$49$\pm$ 5 &SAID~DU13~\protect\cite{du13}\\
		    &  19$\pm$ 2 &$-$38$\pm$ 4 &SAID~CM12~\protect\cite{cm12}\\
                    &  25$\pm$ 5 &$-$49$\pm$ 4 &BnGa12~\protect\cite{bnga12}\\
                    &  66$\pm$18 &$-$223$\pm$29&Kent12~\protect\cite{kent}\\
                    &  18        &$-$28        &MAID~\protect\cite{MAID}\\
                    &  26$\pm$11 &$-$45$\pm$20 &PDG12~\protect\cite{PDG}\\
\hline
\end{tabular}\label{tbl-1}}
\end{table}

\begin{figure*}
\centerline{
\includegraphics[height=11.5cm, angle=90]{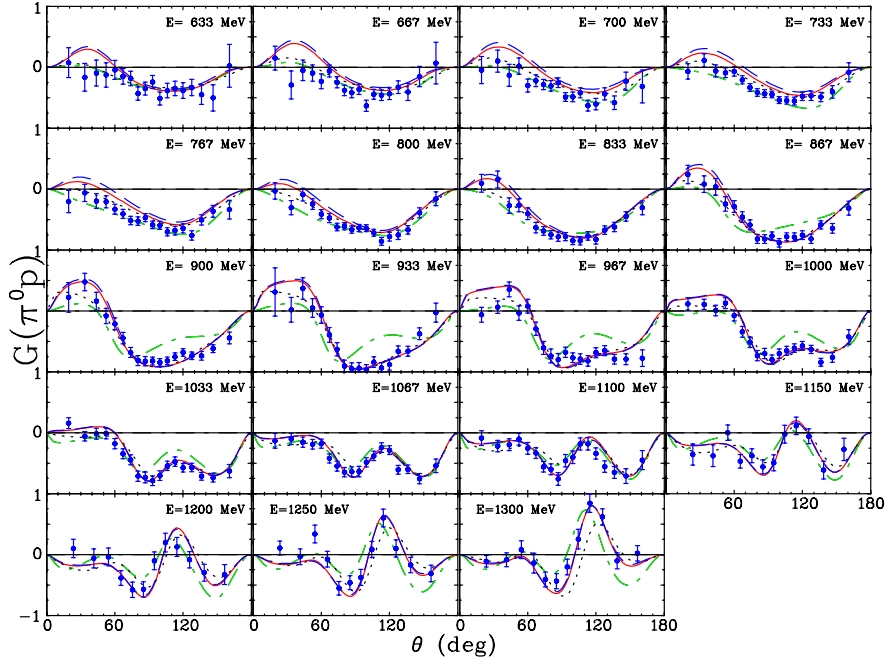}}
\caption{The double-polarization observable $G$ as a
        function of pion production angle in CM. Dashed
        (solid) lines correspond to the SAID (DU13
        \protect\cite{du13} and preliminary solution
        included new CB-ELSA $G$ measuremenents.
        Dotted (dash-dotted) lines give BnGa12
        \protect\cite{bnga12} (MAID07~\protect\cite{MAID}).
        \label{fig-2}}
\end{figure*}

With the inclusion of new high-precision data, our fits are 
becoming more stable and predictive. Plots of recent double
polarized $G$ data, covered E$_\gamma$ = 630 $-$ 1300~MeV 
and $\theta$ = 20 $-$ 160$^\circ$, in Fig.~\ref{fig-2} from 
CB-ELSA~\cite{th12} show that the SAID CM12 fit gives a good 
prediction of this quantity. We have recently analyzed $C_{x'}$ 
(E$_\gamma$ = 460 $-$ 1340~MeV and $\theta$ = 75 $-$ 
140$^\circ$)~\cite{si13} and preliminary $F$ and $T$ data 
(E$_\gamma$ = 440 $-$ 1430~MeV and $\theta$ = 30 $-$ 
160$^\circ$)~\cite{kash13} from Mainz, finding a similarly 
quantitative level of agreement.

\underline{Pion photoproduction off the neutron}. In addition 
to being less precise, experimental data for neutron-target 
photoreactions are much less abundant than those utilizing a 
proton target, constituting only about 15\% of the present 
SAID database~\cite{SAID}.  At low to intermediate energies, 
this lack of neutron-target data is partially compensated by
experiments using pionic beams, e.g., $\pi^-p\to\gamma n$, as 
has been measured, for example, by the Crystal Ball 
Collaboration at BNL~\cite{CB-BNL} for the inverse photon 
energy E$_\gamma$ = 285 $-$ 690~MeV and $\theta$ = 40 $-$
150$^\circ$, where $\theta$ is the inverse production angle 
of pion in the CM frame. This process is free from 
complications associated with the deuteron target. However, 
the disadvantage of using the reaction $\pi^-p\to\gamma n$ 
for the pion photoproduction study is the 5 to 500 times 
larger cross sections for $\pi^-p\to\pi^0n\to\gamma\gamma n$, 
depending on E$_\gamma$ and $\theta$.

\begin{figure}[th]
\centerline{
\includegraphics[height=2.3cm, keepaspectratio]{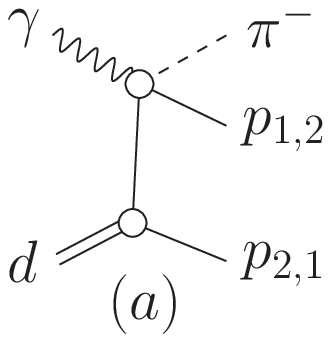}\hfill
\includegraphics[height=2.3cm, keepaspectratio]{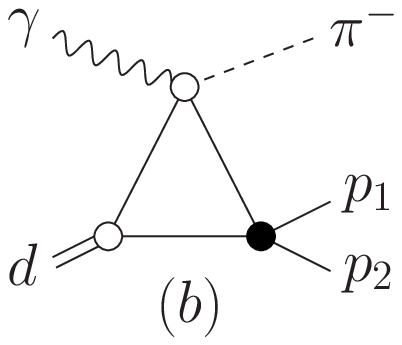}\hfill
\includegraphics[height=2.3cm, keepaspectratio]{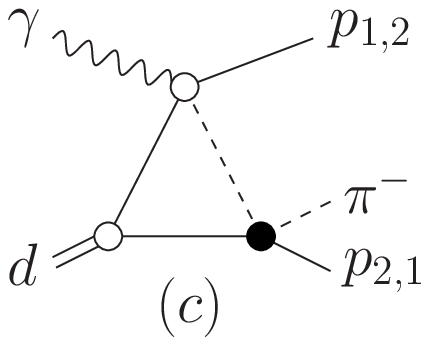}}
\caption{Feynman diagrams for the leading terms of the $\gamma
        d\to\pi^-pp$ amplitude. (a) IA, (b) $pp$-FSI, and (c)
        $\pi$N-FSI. Filled black circles show FSI vertices.
        Wavy, dashed, solid, and double lines correspond to
        the photons, pions, nucleons, and deuterons,
        respectively. \label{fig-3}}
\end{figure}

We extract the $\gamma n\to\pi^-p$ cross section on free 
nucleon from the deuteron data in the quasi-free (QF) 
kinematic region of the $\gamma d\to\pi^-pp$ reaction with 
fast knocked-out proton and slow proton-spectator assumed 
not to be involved in the pion production process. In this, 
so-called impulse approximation (IA)~\cite{CG}, the reaction 
mechanism corresponds to the diagram in Fig.~\ref{fig-3}(a). 
There are 2 critical factors to be taken into account when 
using this approach: (i) the neutron is bound and (ii) there 
are NN- and $\pi N$-FSI effects. 

Item (i) means that the effective mass of the neutron is not 
equal to the mass of the free neutron.  In our former analyses
\cite{gb12,MAMI}, the $\gamma n\to\pi^-p$ amplitude for a 
given E$_\gamma$ and CM pion production angle $\theta$ is 
assumed to be the same as on a free neutron at rest. That is 
why the cross section obtained should be considered as an 
average over energies around E$_\gamma$. The size of the 
averaging region is determined by a smearing of the energy 
owing to the Fermi-motion in the deuteron. The typical scale 
here is 20~MeV in energy.

Item (ii) corresponds to the inclusion of the FSI corrections.
Their leading terms correspond to Feynman diagrams shown on
Fig.~\ref{fig-3}(b,c). Determinations of the $\gamma
d\to\pi^-pp$ differential cross section, with the FSI taken
into account (all the diagrams on Fig.~\ref{fig-3}, were
included) were done recently for the CLAS~\cite{gb12} and
MAMI-B~\cite{MAMI} $\gamma d\to\pi^-pp$ data.  The SAID
phenomenological amplitudes for $\gamma N\to\pi N$~\cite{pionPR},
NN-elastic~\cite{NN}, and $\pi N$-elastic\cite{piN} were used
as inputs to calculate the diagrams in Fig.~\ref{fig-3}. The
Bonn potential~\cite{bonn} was used for the deuteron description.

Recently, we applied our FSI corrections~\cite{FSI} to CLAS 
$\gamma d\to\pi^-pp$ data (E$_\gamma$ = 1050 $-$ 2700~MeV and
$\theta$ = 30 $-$ 160$^\circ$)~\cite{CLAS} to get elementary 
cross sections for $\gamma n\to\pi^-p$~\cite{gb12}.  New CLAS
differential cross sections are quadrupling the world database 
for $\gamma n\to\pi^-p$ above 1~GeV. The FSI correction factor 
for the CLAS kinematics was found to be small, $\Delta\sigma/\sigma 
<10\%$. However, these new cross sections departed significantly 
from our predictions at the higher energies, and greatly modified 
the fit result, which allows to determine new neutron couplings 
(Table~\ref{tbl-2}).

In our recent study~\cite{MAMI}, we addressed to the 
differential cross section measurements for $\gamma n\to
\pi^-p$ in the $\Delta$-isobar region.  The data came from 
MAMI-B (E$_\gamma$ = 300 $-$ 455~MeV and $\theta$ = 60 $-$ 
140$^\circ$)~\cite{GDH}. At energies dominated by the 
$\Delta$-resonance, the isospin I = 3/2 multipoles are 
constrained by extensive studies performed using proton 
targets. The forward peaking structure is due largely to the 
Born contribution, which is well known. As a result, one 
would expect models to give predictions within a tight range.

\begin{table*}
\centering
\caption{Neutron helicity amplitudes $nA_{1/2}$ and $nA_{3/2}$ 
	(in [(GeV)$^{-1/2} \times 10^{-3}$] units).}
\vspace{2mm}
{\begin{tabular}{@{}|cc|ccc|c|@{}}
\hline
Resonance       &$nA_{1/2}$  & Resonance      &$nA_{1/2}$  & $nA_{3/2}$  & Ref. \\
\hline
$N(1535)1/2^-$  &$-$58$\pm$ 6& $N(1520)3/2^-$ &$-$46$\pm$ 6&$-$115$\pm$ 5 &SAID~GB12~\protect\cite{gb12}\\
                &$-$60$\pm$ 3&                &$-$47$\pm$ 2&$-$125$\pm$ 2 &SAID~SN11~\protect\cite{sn11}\\
                &$-$93$\pm$11&                &$-$49$\pm$ 8&$-$113$\pm$12 &BnGa13~\protect\cite{bnga13}\\
		&$-$49$\pm$ 3&		      &$-$38$\pm$ 3&$-$101$\pm$ 4 &Kent12~\protect\cite{kent}\\
                &$-$46$\pm$27&                &$-$59$\pm$ 9&$-$139$\pm$11 &PDG12~\protect\cite{PDG}\\
$N(1650)1/2^-$  &$-$40$\pm$10& $N(1675)5/2^-$ &$-$58$\pm$ 2&$-$80$\pm$ 5  &SAID~GB12~\protect\cite{gb12}\\
                &$-$26$\pm$ 8&                &$-$42$\pm$ 2&$-$60$\pm$ 2  &SAID~SN11~\protect\cite{sn11}\\
                &   25$\pm$20&                &$-$60$\pm$ 7&$-$88$\pm$10  &BnGa13~\protect\cite{bnga13}\\
		&   11$\pm$ 2&		      &$-$40$\pm$ 4&$-$68$\pm$ 4  &Kent12~\protect\cite{kent}\\
                &$-$15$\pm$21&                &$-$43$\pm$12&$-$58$\pm$13  &PDG12~\protect\cite{PDG}\\
$N(1440)1/2^+$  &   48$\pm$ 4& $N(1680)5/2^+$ &   26$\pm$ 4&$-$29$\pm$ 2  &SAID~GB12~\protect\cite{gb12}\\
                &   45$\pm$15&                &   50$\pm$ 4&$-$47$\pm$ 2  &SAID~SN11~\protect\cite{sn11}\\
                &   43$\pm$12&                &   34$\pm$ 6&$-$44$\pm$ 9  &BnGa13~\protect\cite{bnga13}\\
		&   40$\pm$ 5&		      &   29$\pm$ 2&$-$59$\pm$ 2  &Kent12~\protect\cite{kent}\\
                &   40$\pm$10&                &   29$\pm$10&$-$33$\pm$ 9  &PDG12~\protect\cite{PDG}\\
\hline
\end{tabular} \label{tbl-2}}
\end{table*}

We have included the new neutron cross sections from the 
CLAS and MAMI-B experiments in a number of multipole 
analyses covering incident photon energies up to 2.7~GeV, 
using the full SAID database~\cite{SAID}, in order to 
gauge the influence of these measurements, as well as 
their compatibility with previous experiments. The 
solution, GB12~\cite{gb12}, uses the same fitting form as 
our recent SN11 solution~\cite{sn11}.  A second fit, GZ12, 
instead used the recently proposed form based on a unified 
Chew-Mandelstam parametrization of the GW DAC fits to 
both $\pi$N elastic scattering and photoproduction
\cite{cm12}.

Table~\ref{tbl-2} shows that the new SAID GB12 $nA_{1/2}$ 
and $nA_{3/2}$ helicities sometimes have a significant 
deviation from the previous SAID SN11~\cite{sn11} 
determination and PDG12~\cite{PDG} values, e.g., for 
$N(1650)1/2^-$, $N(1675)5/2^-$, and $N(1680)5/2^+$. 
While BnGa13 group~\cite{bnga13} used the same (almost) 
data to fit them as we are while BnGa13 has several new 
ad hoc resonances.  Meanwhile, BnGa13 determination is 
different for $N(1535)1/2^-$, $N(1650)1/2^-$, and 
$N(1680)5/2^+$.

\underline{Summary}. Future progress in the database 
development is expecting from tagged-photon fasilities as 
JLab, MAMI-C, SPring-8, CB-ELSA, and ELPH.  Partial-wave 
analyses will clearly benefit from the constraints provided 
by these new data, which highlight the importance of new 
polarization observables in providing a stringent test of 
PWA, even in kinematic regions where a large number of 
cross section and polarization observables are already 
present in the world database. An accurate PWA must 
ultimately describe a complete set of observables. The 
current data and future experiments exploiting these 
polarimetry developments at large acceptance detectors 
will be a key part to achieving this complete measurement. 

In this regard, future experiments to measure unpolarized
and the spin polarization of neutrons are already planned 
at MAMI-C.  Measurements of such observables with large 
acceptance are crucial to the world program aiming to 
determine the excitation spectrum of the nucleon.

We proposed to perform a precision measurement of 
$d\sigma/d\Omega$ in the reactions $\gamma d\to\pi^-pp$ and
$\gamma d\to\pi^0np$ in the tagged-photon energy region from 
threshold to 800~MeV~\cite{mami12} and then to 1500~MeV
\cite{mami13}.  The $d\sigma/d\Omega$ for the processes 
$\gamma p\to\pi^-p$ and $\gamma p\to\pi^0n$ will be extracted 
from these CB@MAMI-C measurements accounting for Fermi motion 
effects in IA~\cite{CG} as well as NN- and $\pi$N-FSI effects 
beyond the IA. Data below 800~MeV were taken in March of 2013 
and analysis is in progress.  Consequential calculations of 
the FSI corrections, as developed by our GW-ITEP Collaboration, 
will be applied.  We will extend our FSI code~\cite{FSI} to 
extract $\gamma n\to\pi^0n$ data from $\gamma d\to\pi^0np$ 
measurements as well.  Polarized measurements will help to 
bring more physics in. FSI corrections need to apply.

\underline{Acknowledgments}. This work was supported in part 
by the U.~S. DOE Grant No. DE-FG02-99ER41110. 


\end{document}